\documentclass[%
 reprint,
 amsmath,amssymb,
 aps,
]{revtex4-2}

\usepackage{graphicx}
\usepackage{dcolumn}
\usepackage{bm}
\usepackage[version=4]{mhchem}

\begin{document}

\preprint{APS/123-QED}

\title{Understanding the Stability for $LiNi_{0.5}Mn_{0.5}O_2$ as a Co-free positive electrode material}

\author{Xinyi Liu}
\author{Dongyan Zhang}%
  \email{zhangdongyan@xidian.edu.cn}
 \author{Zhimin Li}
 \email{zmli@mail.xidian.edu.cn}
 \author{Maolin Zhang}
 \author{Yangxi Yan}
\affiliation{%
 School of Advanced Materials and Nanotechnology, Xidian University, Xi’an 710126, China
}%

\author{PangPang Wang}
\affiliation{
 Nanomaterials Group, Institute of System, Information Technologies and Nanotechnologies (ISIT), Fukuoka Industry-Academia Symphonicity (FiaS), 4-1 Kyudaishinmachi, Nishi-ku, Fukuoka 819-0388, Japan
}%

\author{Ri-ichi Murakami}
\affiliation{%
 School of Mechanical Engineering, Chengdu University, Chengdu 610106, P. R. China
}%

\date{\today}

\begin{abstract}
For understanding the stability of Co-free positive electrode material, $LiNi_{0.5}Mn_{0.5}O_2$ was synthesized with different addition amount of lithium during calcination. The valence states of transition metal in the prepared samples were determined by combining accurate stoichiometry analysis via ICP, magnetic moment measurement via SQUID, and element valence analysis via XPS with Ar ion etching. Unexpectedly, the transition metals, Ni and Mn,  at interior and surface of  $LiNi_{0.5}Mn_{0.5}O_2$ particles show different electrochemical properties. This answer lingering questions of Li de-intercalation mechanism in  $LiNi_{0.5}Mn_{0.5}O_2$.

\end{abstract}

\maketitle

\section{\label{sec:Intro}Introduction}

Layered lithium nickel manganese oxides ($LiNi_xMn_{1-x}O_2$) are promising alternative positive electrode materials to the currently commercial $LiCoO_2$ and Ternary materials for lithium ion batteries (LIBs) \cite{HU2020137127, Voronina2020}. $LiNi_xMn_{1-x}O_2$ cathode materials are expected to possess as high theoretical capacity as $LiCoO_2$ of approximately 279 $mAh\cdot g^{-1}$ \cite{MENG2008489}, which is based on assumptions that \cite{ArcangeloCeleste2021ACSAEM,Won-SubYoon2003ChemMater,Deb2006JAP,ABDELGHANY2007EA} (1). Nickel and manganese are $Ni^{2+}$,$Mn^{4+}$ in $LiNi_xMn_{1-x}O_2$, (2). The $Ni^{2+}/Ni^{4+}$ redox pairs contribute to electrochemical activity while all lithium ions are extracted, and $Mn^{4+}$ ions remain as $Mn^{4+}$ throughout the cycling process. However, several issues still doubt researchers including diverse reported specific capacities \cite{MANIKANDAN2011PowerSources,ZHANG2008PowerSources,DARBAR2020PowerSources,YANG2017JALCOM,HU2013121}, capacity loss during the initial cycles \cite{LiHong2015RareMet,Zhang2015Ionics,HU2013121}, and poor lithium ion diffusivity \cite{Yabuuchi_2010,ABDELGHANY2019100207}. A lot of efforts have been made to elucidate these problems and to improve the electrochemical properties based on the above mentioned assumptions \cite{MANIKANDAN2011PowerSources,ZHANG2008PowerSources,DARBAR2020PowerSources,YANG2017JALCOM,HU2013121,LiHong2015RareMet,Zhang2015Ionics,HU2013121,Yabuuchi_2010,ABDELGHANY2019100207}. For example, the capacity loss during the initial cycles was considered to be arising from surface reaction of cathode with electrolyte, so various coating methods have been developed to isolate the cathode from electrolyte \cite{ABDELGHANY2019100207}; the poor rate capacity was consider to mainly be attributed to cation mixing at lithium site, so a strategy has been widely used that lattice was strained by ions doping to inhibit nickel occupying the lithium site \cite{Jia2020Ionics}. This theory about $Ni^{2+}$ and $Mn^{4+}$ in $LiNi_xMn_{1-x}O_2$ originated from the conventional element valence analysis characterized by X-ray photoelectron spectroscopy (XPS)\cite{Quinlan_2013} and cyclic voltammetry \cite{PERIASAMY20061360}. However,  XPS could only characterize properties originating from surface several manometers \cite{AZIZ201781}, and redox potentials of $Ni^{2+}/Ni^{3+}$ and $Ni^{3+}/Ni^{4+}$ are close \cite{PERIASAMY20061360}. Hence, the deduction of $Ni^{2+}$ and $Mn^{4+}$ in $LiNi_xMn_{1-x}O_2$ is a fantasied edifice, and a lot of doubts hangs over thousands of researchers to bring the cobalt free layered lithium nickel manganese oxides into practical applications. The urgent task is to confirm the chemical structure of $LiNi_xMn_{1-x}O_2$.

Herein, valence states of transition metals in as-prepared cobalt free layered lithium nickel manganese oxides with nominal stoichiometry of $LiNi_{0.5}Mn_{0.5}O_2$ were determined by combining  accurate stoichiometry analysis via ICP,  magnetic moment measurement via SQUID, and depth-profiling valence via XPS. Unexpectedly, the transition metals, Ni and Mn,  at interior and surface of  $LiNi_{0.5}Mn_{0.5}O_2$ particles demonstrated different electrochemical properties. Ni at the surface of $LiNi_{0.5}Mn_{0.5}O_2$ was +2, but at interior was +3. Mn at the surface was +4, at interior was +3.  XPS for the Mn 3s orbital with Ar ion etching $LiNi_{0.5}Mn_{0.5}O_2$ particles confirmed that binding energy presented significant difference for surface and interior. In the following, we should see such an inhomogeneous valence of transition metal indeed existed in as-prepared cobalt free layered lithium nickel manganese oxides,  $LiNi_{0.5}Mn_{0.5}O_2$,  and its influence on the stability and electrochemical properties.

\section{\label{sec:expe}Experimental}
The cobalt free layered lithium nickel manganese oxide,  $LiNi_{0.5}Mn_{0.5}O_2$,  were prepared by sol-gel method using lithium acetate ($LiCH_3COO \cdot 2H_2O$, MACKLIN, 99.9\%), nickel acetate ($Ni(CH_3COO)_2 \cdot 4H_2O$, Shanghai Chemical Reagent Research Institute, 98\%),  and manganese acetate ($Mn(CH_3COO)_2 \cdot 4H_2O$, Shanghai Chemical Reagent Research Institute, 90\%) as source, citric acid as complexing agent. The samples were calcined at 900 ℃ for 16 h  to obtain the final products after preparing the xerogel powders as procedures of dissolving metal salts in deionized water, adding citric acid, heating in water bath at 80℃ for 4 hours to obtain sol, drying in vacuum drying oven at 120℃ for 12 h to obtain xerogel, and grinding to powders. Due to the volatilization of lithium during calcination, 10\%, 15\%, and 20\% excess $LiCH_3COO \cdot 2H_2O$ were added during the initial dissolution procedure, and the obtained products were denoted as L10, L15, and L20. 

The crystal structure of prepared $LiNi_{0.5}Mn_{0.5}O_2$ samples was analyzed by X-ray diffractometry (XRD) using a Bruker D8 Advance apparatus equipped with a Cu K$_{\alpha}$ X-ray source ($\lambda = 1.5406 $ \AA). XRD measurements were collected in the $2\theta$ range $10-80 ^{\circ}$. The stoichiometry of each sample was decided by inductively coupled plasma-optical emission spectrometry (ICP-OES) using Agilent 5100. Magnetic measurements for susceptibility were carried out with with a fully automated magnetometer (MPMS-3 from Quantum Design) using a ultra-sensitive Superconducting Quantum Interface Device (SQUID) in the temperature range of 3-300 K. Chemical state identifications were made using XPS (Thermo Scientific K-Alpha$^+$) with a monochromatic Al $K_\alpha$ source. The XPS data were collected before and after samples were etched by Ar ions with the depths of 20 nm and 40 nm.
The electrochemical properties of the samples were tested at room temperature in cells with metallic lithium as anode electrode. Cell assembly and measurements were carried out following the experimental procedure previously described \cite{LiJieJMCA2021J} using Land CT2991A laboratory-cell test system (Wuhan Jinnuo Electric Co. Ltd., China) . Voltage curves were recorded at  0.05 C rate for initial charge-discharge cycle, and at 1C for the following capacity retention test. 

\section{\label{sec:Results}Results and Discussions}
\begin{figure}
\includegraphics[width=0.8\linewidth]{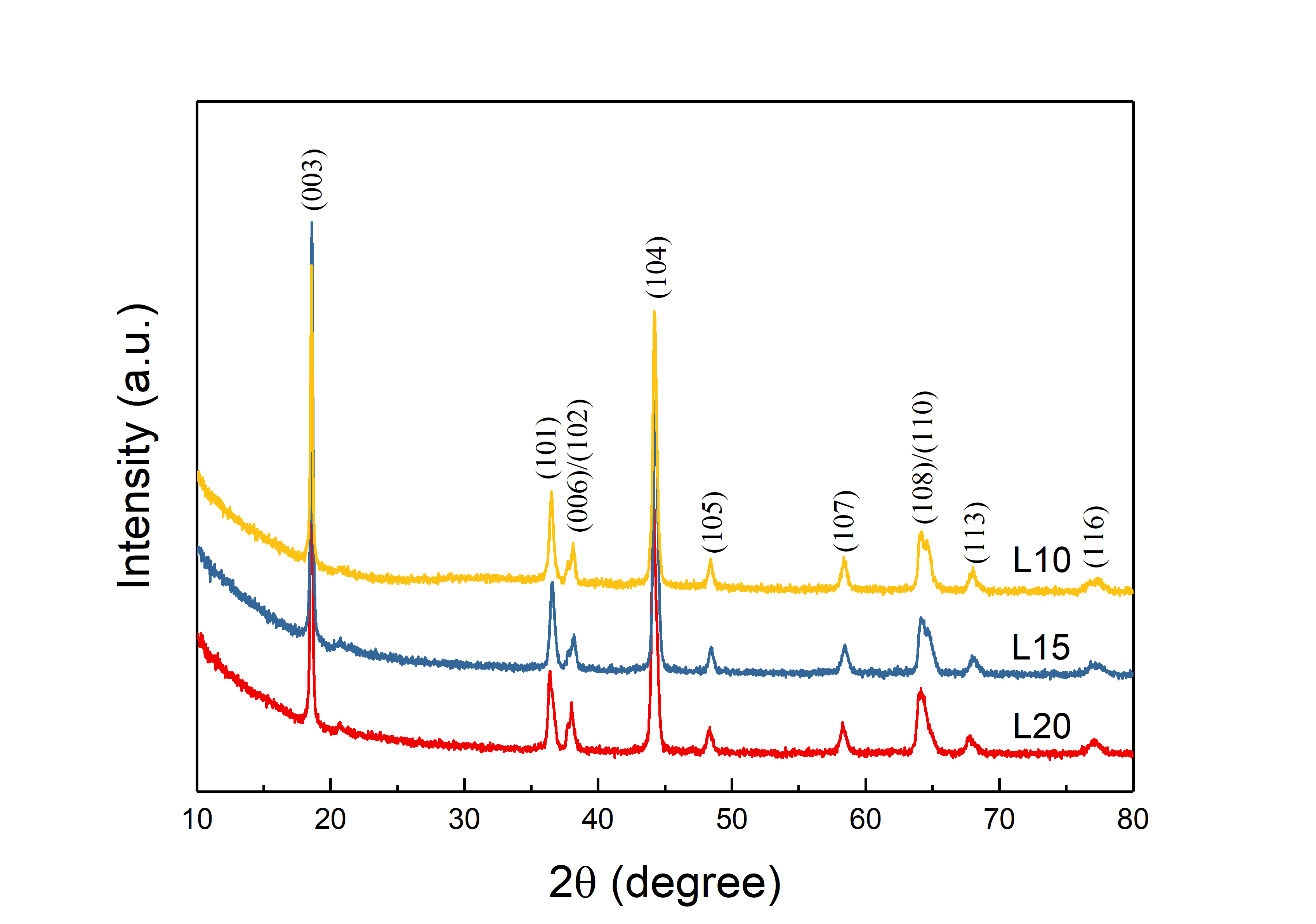}
\caption{\label{fig:XRD} XRD patterns of Samples L10, L15, and L20.}
\end{figure}

Figure \ref{fig:XRD} shows the XRD patterns of as-prepared Samples L10, L15, and L20. All patterns were indexed by a typical $\alpha-NaFeO_2$ hexagonal layered structure. The patterns for all samples did not display remarkable difference, indicating that the addition amount of excess lithium did not change the crystalline profile. However, some weak diffraction peaks between 20$^o$ and 25$^o$ were observed in XRD patterns of Samples L15, and L20. These are arising from the superlattice order of Li-Mn ions in transition-metal layers, demonstrating the existence of Li-rich $Li_2MnO_3$ phase. The existence of $Li_2MnO_3$ in Samples L15 and L20 was attributed to the massive overdose of $LiCH_3COO \cdot 2H_2O$ during synthesis process. The accurate stoichiometry of each sample was determined by ICP-OES.

Table \ref{tab:ICP} shows the stoichiometry and valence of cations in as-prepared samples decided by ICP-OES and SQUID. It demonstrates that the Ni/Mn mole ratio is close to the nominal value of 1:1, however due to the occupation of excess lithium at transition metal sites,  the stoichiometry of Sample L15 and L20 deviate from the designed chemical formula, $LiNi_{0.5}Mn_{0.5}O_2$. Among the as-prepared samples, L10 approximately concurred with the design, but L15 and L20 possess excess lithium at transition metal site  and the concentrations of Mn are little higher than that of Ni. 

\begin{table*}
\caption{\label{tab:ICP}The stoichiometry and valence for each element in each sample.},
\begin{ruledtabular}
\begin{tabular}{cccccc}
Sample & stoichiometry by ICP & $\theta_p$ &$C_p$ & $\mu_{eff}$ &stoichiometry with Valence\\ \hline
L10 &$Li(Li_{0.01}Ni_{0.49}Mn_{0.5})O_2$ & -123 K & 2.30 $emu \cdot K/mol$ & 4.32 $\mu_B$ & $Li(Li_{0.01}^{1+}Ni_{0.49}^{3+}Mn_{0.48}^{3+}Mn_{0.02}^{4+})O_2 $ \\
L15 &$Li(Li_{0.1}Ni_{0.42}Mn_{0.48})O_2$ & -109 K& 1.71  $emu \cdot K/mol$ & 3.71 $\mu_B$ & $Li(Li_{0.1}^{1+}Ni_{0.03}^{2+}Ni_{0.39}^{3+}Mn_{0.25}^{3+}Mn_{0.23}^{4+})O_2 $ \\
L20 &$Li(Li_{0.1}Ni_{0.44}Mn_{0.46})O_2$ & -122 K& 1.72  $emu \cdot K/mol$ & 3.72 $\mu_B$ & $Li(Li_{0.1}^{1+}Ni_{0.01}^{2+}Ni_{0.43}^{3+}Mn_{0.25}^{3+}Mn_{0.21}^{4+})O_2 $ \\
\end{tabular}
\end{ruledtabular}
\end{table*}

\begin{figure}
\includegraphics[width=0.8\linewidth]{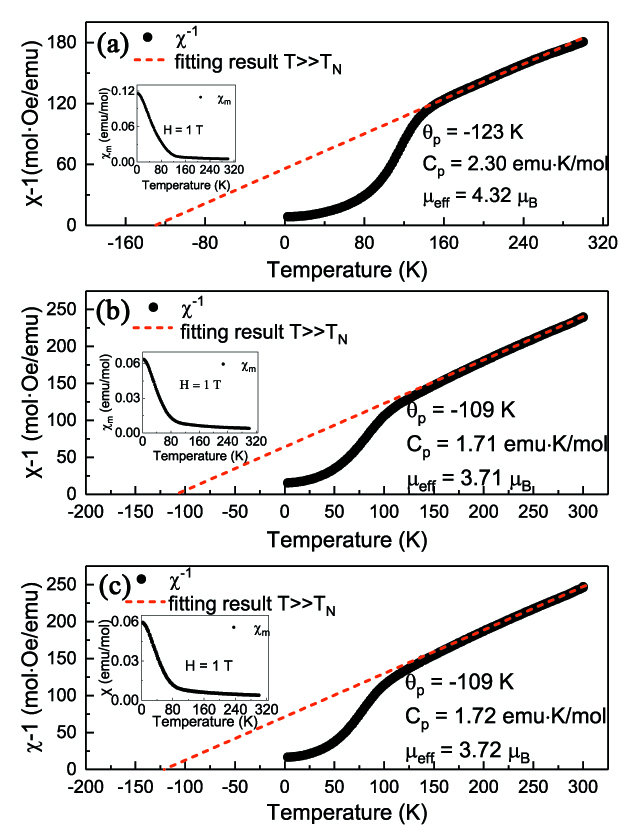}
\caption{\label{fig:SQUID} Temperature dependence of the reciprocal magnetic susceptibility for (a)Sample Li10, (b)Sample Li15, (c)Sample Li20. Insert shows the FC temperature dependence of the magnetic susceptibility.}
\end{figure}

Because lithium possesses one $2s$ orbital electron, either lithium ions located at Li sites or transition metal sites  should be valence of +1. Owing to the condition of electroneutrality, the average valence of cations located at transition metal sites should be +3. This indicates that nickel and manganese ions located at transition metal sites in $LiNi_{0.5}Mn_{0.5}O_2$ should be the pair of either $Ni^{2+}/Mn^{4+}$ or $Ni^{3+}/Mn^{3+}$. Both nickel and manganese ions possess unpaired $3d$ orbital electrons resulting in magnetic moment, and nickel or manganese ions with different valence contributes different magnetic moments. Hence, the valence of nickel and manganese ions could be deduced from magnetic measurement data. Figure \ref{fig:SQUID} shows the temperature dependence of the reciprocal magnetic susceptibility for Sample L10, L15, and L20. Above 200 K, the samples start to present a paramagnetic (PM) behavior which follows the Curie–Weiss law,
\begin{equation}
\chi = \frac{C_p}{T-\theta_p}
\end{equation}
where $C_p$ is the Curie constant and $\theta_p$ is the Weiss temperature. For temperature much higher than Curie temperature where samples behave like paramagnetic materials, the linear fit of the temperature dependence of the reciprocal magnetic susceptibility  gives a Weiss constant $\theta_p$  and a Curie constant $C_p$.  The fitted results are shown in Figure \ref{fig:SQUID} and Table \ref{tab:ICP}. The negative $\theta_p$ value suggests that antiferromagnetic interactions are present in obtained $LiNi_{0.5}Mn_{0.5}O_2$. After the Curie constant, $C_p$, is determined experimentally by fitting the linear $\chi^{-1}(T)$ curve in the paramagnetic region, one can estimate the experimental value of effective moment $\mu_{eff}$  and then know the electronic configuration of the magnetic cation using the following equation\cite{Julien2006Magnetic},
\begin{equation}
\mu_{eff}  = \sqrt{\frac{3k_BC_p}{N_A\mu_B^2}} = 2.84\sqrt{C_p}
\end{equation}
where  $N_A$ is the molar concentration of ions, $\mu_B$ the Bohr magneton, $k_B$ the Boltzmann constant. The calculated effective moment $\mu_{eff}$ are listed in Table \ref{tab:ICP}. Combining the stoichiometry data by ICP-OES, the stoichiometry with valence for each element could be obtained according to the conservation of mass (Eqs. \ref{eq:NiBlan} and \ref{eq:MnBlan}), charge (Eq. \ref{eq:chargeBlan}) and moment (Eq. \ref{eq:spinBlan}), 
\begin{subequations}
\begin{eqnarray}
\sum n_{Ni^{i+}} &=& n_{Ni} \label{eq:NiBlan}
\\
\sum n_{Mn^{i+}} &=& n_{Mn} \label{eq:MnBlan}
\\
\sum i*n_{M^{i+}} &=& 3 \label{eq:chargeBlan}
\\
\sum\mu_{M^{i+}} &=& \mu_{eff} \label{eq:spinBlan}
\end{eqnarray}
\end{subequations}
where $n$ is the concentration of each element, $i$ represents the valence, $M$ refers to the ions at transition metal site, $\mu$ indicates the magnetic moment.  Sample L10 whose stoichiometry approximately concurred with the design possessed mainly $Ni^{3+}$ and $Mn^{3+}$, while Samples L15 and L20 possess $Ni^{2+}$ and $Mn^{4+}$ besides of $Ni^{3+}$ and $Mn^{3+}$. 

\begin{figure}
\includegraphics[width=0.8\linewidth]{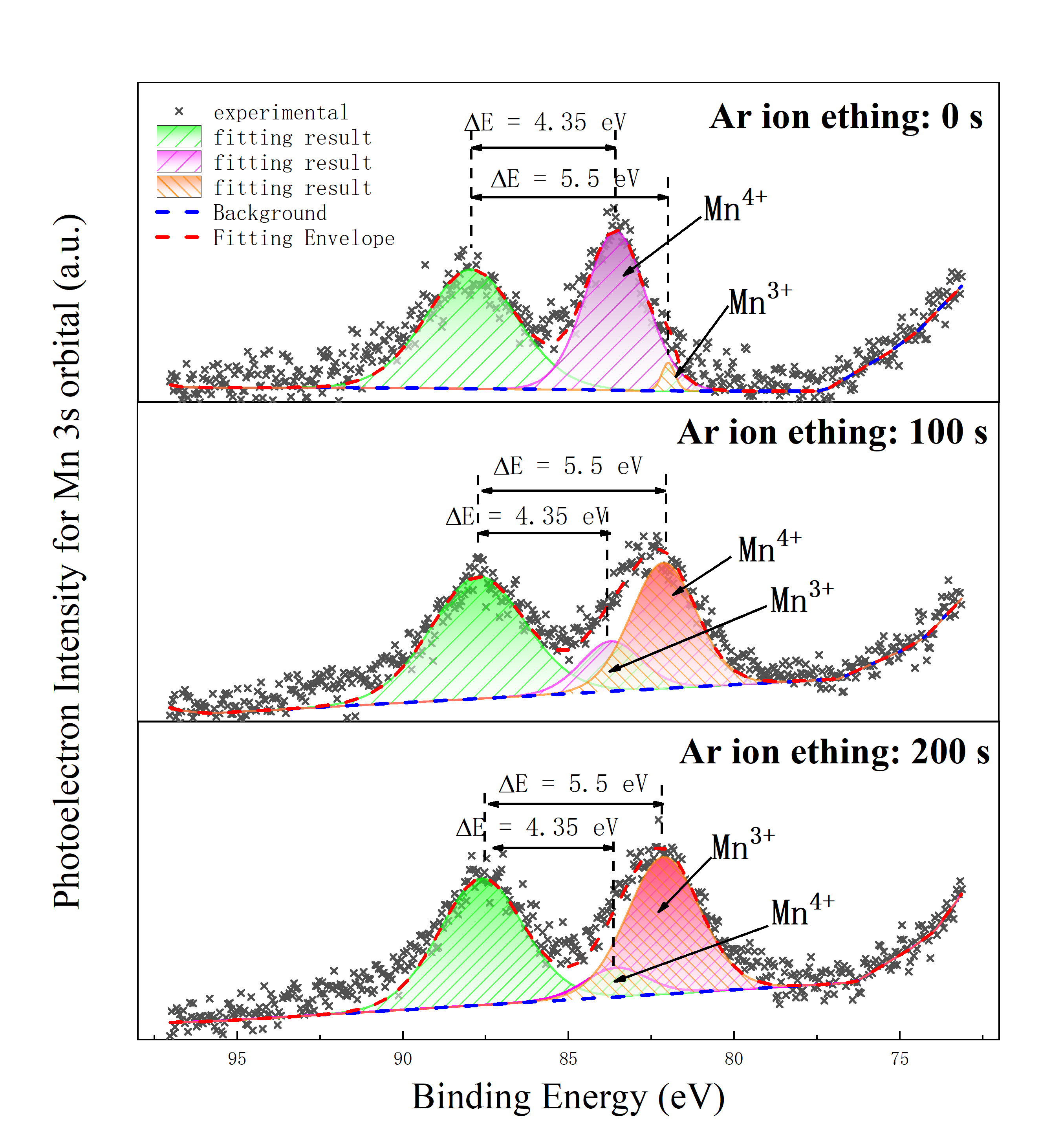}
\caption{\label{fig:XPS_Mn3s} XPS of $Mn$ 3s orbital for Sample L10 with different Ar ion etching depth.}
\end{figure}

Different from the widely reported that $Mn$ in $LiNi_{0.5}Mn_{0.5}O_2$ is of $Mn^{4+}$ to form the pair of $Ni^{2+}/Mn^{4+}$, the combination of ICP-OES and SQUID measurements demonstrate that the major of $Mn$ in $LiNi_{0.5}Mn_{0.5}O_2$ is of $Mn^{3+}$.  This difference is arising from the probing depth between SQUID and XPS. Magnetic measurement via SQUID collect the signal originating from  the entire sample, but XPS probe the surface profile of the sample.  Similar with reported in literature, the XPS  data collected from the natural surface of sample L10 demonstrate $Mn$ cations is mainly consist of $Mn^{4+}$ according to the XPS of Mn $3s$ orbital as shown in Figure \ref{fig:XPS_Mn3s}. Combining with Ar ion etching, depth-profiling XPS analysis could be enabled. Herein, Ar ions were employed to etch the surface with 100 and 200 seconds into the depths of 20 nm and 40 nm. As the etching depth increases, the concentration of $Mn^{3+}$ grows confirmed by the increasing in the energy difference between the multiplet  states from 4.35 eV ($Mn^{4+}$) to 5.5 eV ($Mn^{3+}$) presented in Figure \ref{fig:XPS_Mn3s}. Thus, we could conclude that the $Mn^{4+}$ distributed at the surface of layered lithium nickel manganese oxides to form $LiNi_{0.5}^{2+}Mn_{0.5}^{4+}O_2$ and $Mn^{3+}$ centered the interior to form $LiNi_{0.5}^{3+}Mn_{0.5}^{3+}O_2$. 

\begin{figure}
\includegraphics[width=1.0\linewidth]{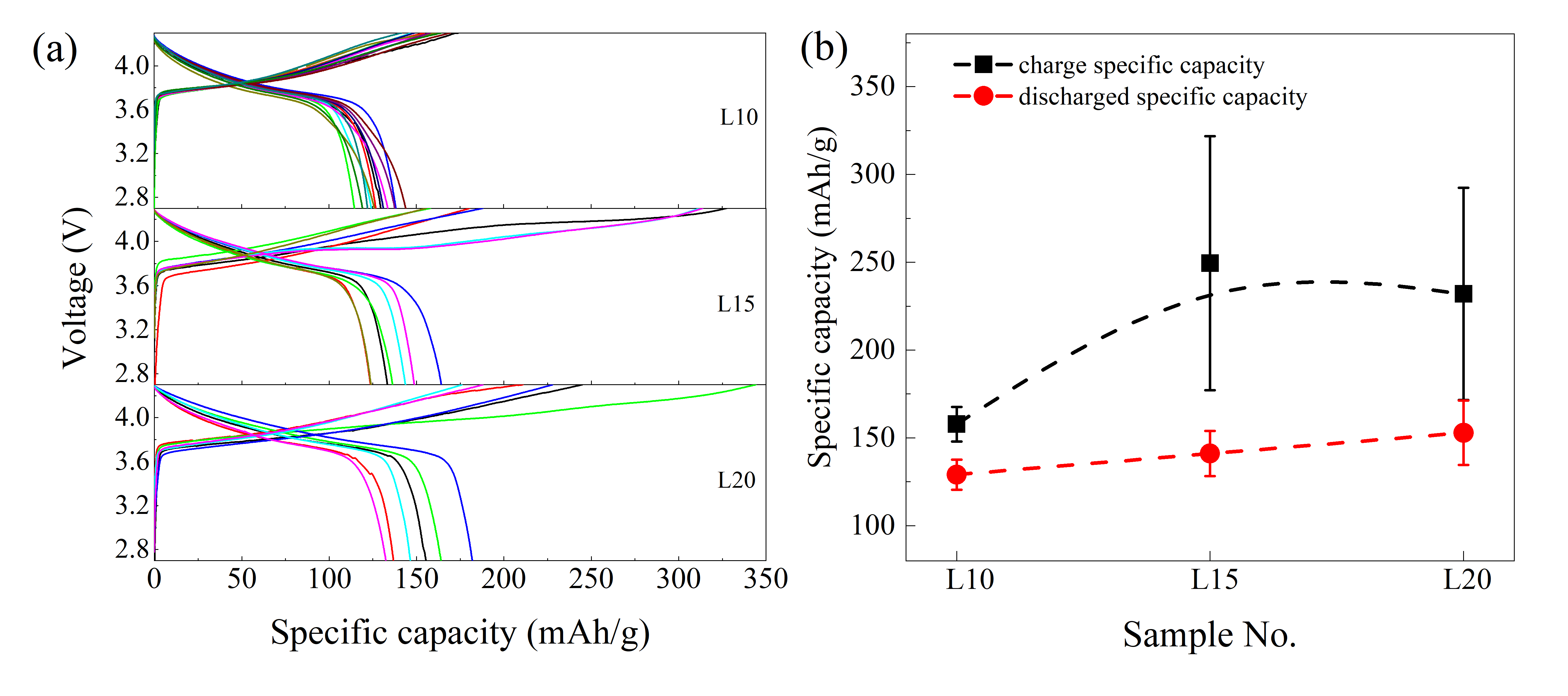}
\caption{\label{fig:capacity} (a)Charge/discharge specific capacity vs. voltage curves of assembled cells with L10, L15 and L20 cathode materials, (b)Statistical charge/discharge specific capacity for assembled cells with L10, L15, and L20 cathode materials.}
\end{figure}

Because surface and interior of $LiNi_{0.5}Mn_{0.5}O_2$ behave different chemistry states, the delithiation rely on different reodx thermodynamics. For surface
\begin{equation}
\begin{aligned}
& LiNi_{0.5}^{2+}Mn_{0.5}^{4+}O_2 \\
&  \ce{->} Li^++e^-+Ni_{0.5}^{4+}Mn_{0.5}^{4+}O_2 \\
\end{aligned}
\end{equation}
For interior
\begin{equation}
\begin{aligned}
& LiNi_{0.5}^{3+}Mn_{0.5}^{3+}O_2 \\
&  \ce{->} 0.5Li^++0.5e^-+Li_{0.5}Ni_{0.5}^{4+}Mn_{0.5}^{3+}O_2 \\
\end{aligned}
\end{equation}
Due to the domination of $Ni^{3+}$/$Mn^{3+}$ in $LiNi_{0.5}Mn_{0.5}O_2$, nearly half of the theoretical specific capacity could be delivered. During the lithiation, $Ni^{4+}$/$Ni^{3+}$ redox occurred. Hence, existence of $Ni^{2+}$ led to initial capacity loss. If $Mn^{4+}$ concentration increased, the coulombic efficiency during initial charge/discharge cycle would be reduced. When the addition of lithium increase, excess $Li^+$ ions would occupied transition metal sites, resulting in double amount of manganese behave $Mn^{4+}$ to balance the valence for an average of $+3$ at transition metal sites. The induced  $Mn^{4+}$ possesses more neighbour transition metal sites, resulting in some $Ni^{2+}$ to achieve the local electric neutrality. As shown in Table \ref{tab:ICP}, and $Ni^{2+}$ ions were identified in Sample L15 and L20. Due to the existence of $Ni^{2+}$ and $Li^+$ at transition metal sites, the coulombic efficiency would be terrible during initial charge/discharge cycling. Figure \ref{fig:capacity} shows the charge/discharge profiles of assembled cells with L10, L15 and L20 cathode materials. As above discussed, assembled cells with Sample L10 cathode materials delivered average 130 mAh/g which is nearly half of the theoretical value of 279 mAh/g (assuming all lithium ions were extracted). When excess lithium were added, lithium-rich manganese oxides generated and excessive capacity was charged as shown in Figure \ref{fig:capacity}a (L15 and L20). However, the excessive charged capacity was irreversible, and the discharged capacity reverted to the level of half theoretical value depending on the redox of $Ni^{3+}$ (Figure \ref{fig:capacity}).

\section{\label{sec:Con}Conclusion}
$LiNi_{0.5}Mn_{0.5}O_2$ with different addition amount of lithium was synthesized. The stoichiometry and valence states of transition-metal cations in as-prepared samples were analyzed by the combination of ICP-OES and SQUID measurements, which demonstrate the domination of $Ni^{3+}$/$Mn^{3+}$ in as-prepared $LiNi_{0.5}Mn_{0.5}O_2$. XPS equipped with Ar ion etching was employed to analyze the distribution of $Mn$ with different valence in the samples, which demonstrate that $Mn^{4+}$  lay at surface and $Mn^{3+}$ centered at the interior. This answer lingering questions of Li de-intercalation mechanism in  $LiNi_{0.5}Mn_{0.5}O_2$ that half of the theoretical capacity was arising from the $Ni^{3+}$/$Ni^{4+}$ redox, excess capacity from existence of lithium-rich manganese oxide, and capacity loss during the initial charge/discharge cycling owing to the existence of $Ni^{2+}$.
\section*{Acknowledgement}
This work was supported by the National Natural Science Foundation of China (Grant No. 52173227, 61974114), and the Fundamental Research Funds for the Central Universities (Grand No. JB211404).

\nocite{*}
\bibliography{apssamp}
\end{document}